\documentclass{kluwer}
\usepackage{graphicx}
\newcommand{\vek}[1]{\bf{#1}}
\newcommand{\bea}{\begin{eqnarray}}
\newcommand{\eea}{\end{eqnarray}}
\newcommand{\nn}{\nonumber}

\newcommand{\aap}{Astronomy \& Astrophysics~}
\newcommand{\apj}{Astrophysical Journal~}
\newcommand{\aj}{Astronomical Journal~}

\newcommand{\plotone}[1]{\includegraphics[width=5.0in,height=3.in]{#1}}
\newcommand{\comment}[1]{ {} }
\begin{document}
\begin{article}
\begin{opening}
\title{Measuring Black Hole Spin in OJ287}
\author{Valtonen$^{1,8}$, S.~Mikkola$^{1}$, H.~J.~Lehto$^{1}$, T.~Hyv\"onen$^{1}$, K.~Nilsson$^{1}$, D.~Merritt$^2$, A.~Gopakumar $^{3}$, H.~Rampadarath$^{4,5}$, R.~Hudec$^{6,7}$,
M.~Basta$^{6}$ and R.~Saunders$^{5}$}
\institute{$^1$Tuorla Observatory, Department of Physics and Astronomy, University of Turku,
    21500 Piikki\"o, Finland\\
$^2$ Centre for Computational Relativity and Gravitation, Rochester Institute of Technology,
78 Lomb Memorial Drive, Rochester, NY 14623, USA \\
$^3$Tata Institute of Fundamental Research, Mumbai 400005, India\\
$^4$School of Physics and Astronomy, University of Manchester, Alan Turing Building, Oxford Road, Manchester M13 9PL, UK\\
$^5$ Department of Physics, University of the West Indies, St. Augustine, Trinidad \& Tobago\\
$^6$ Astronomical Institute, Academy of Sciences, Fricova 298, 25165 Ondrejov, Czech Republic\\
$^{7}$ Czech Technical University in Prague, Faculty of Electrical
Engineering, TechnickÃ¡ 2, 166 27 Praha 6, Czech Republic\\
$^8$ Helsinki Institute of Physics, PL 64, FIN-00014 Helsingin yliopisto, Finland
}

\runningauthor{Valtonen et al.}
\runningtitle{Black hole spin}
\begin{abstract}

We model the binary black hole system OJ287 as a spinning primary and a non-spinning secondary. It is assumed that the primary has an accretion disk which is impacted by the secondary at specific times. These times are identified as major outbursts in the light curve of OJ287. This identification allows an exact solution of the orbit, with very tight error limits. Nine outbursts from both the historical photographic records as well as from recent photometric measurements have been used as fixed points of the solution: 1913, 1947, 1957, 1973, 1983, 1984, 1995, 2005 and 2007 outbursts. This allows the determination of eight parameters of the orbit. Most interesting of these are the primary mass of  $1.84\cdot 10^{10} M_\odot$, the secondary mass  $1.46\cdot 10^{8} M_\odot$, major axis precession rate  $39^\circ\!.1$ per period, and the eccentricity of the orbit $0.70$. The dimensionless spin parameter is  $0.28\:\pm\:0.01$ (1 sigma). The last parameter will be more tightly constrained in 2015 when the next outburst is due. The outburst should begin on 15 December 2015 if the spin value is in the middle of this range, on 3 January 2016 if the spin is 0.25, and on 26 November 2015 if the spin is 0.31. We have also tested the possibility that the quadrupole term in the Post Newtonian equations of motion does not exactly follow Einstein's theory: a parameter $q$ is introduced as one of the 8 parameters. Its value is within 30\% (1 sigma) of the Einstein's value $q = 1$. This supports the $no-hair~theorem$ of black holes within the achievable precision. We have also measured the loss of orbital energy due to gravitational waves. The loss rate is found to agree with Einstein's value with the accuracy of 2\% (1 sigma). There is a possibility of improving the accuracy of both quantities using the exact timing of the outburst on 21 July 2019. Because of closeness of OJ287 to the Sun ($8 - 12^{\circ}$), the observations would be best carried out by a telescope in space.
\end{abstract}

\end{opening}
\section{Introduction}
The range of celestial mechanics has been expanding from purely Solar system studies to binary stars and planetary systems around other stars, and in recent years also to binary pulsars. The most recent application of celestial mechanics is in the binary system of two supermassive black holes in the quasar OJ287. The identification of this system as a likely binary was made as early as 1982 (Sillanp\"a\"a et al. 1988), but since the mean period of the system is as long as 12 yr, it has taken a quarter of century to find convincing proof that we are indeed dealing with a binary system (Valtonen et al. 2008). The primary evidence for a binary system comes from the optical light curve. By good fortune, the quasar OJ287 has been photographed accidentally since 1890's, well before its discover in 1968 as an extragalactic object. The light curve of over hundred years (Figure 1) shows a pair of outbursts at 11 - 14 yr intervals and the two brightness peaks are separated by 1 - 3 yrs.
\begin{figure}[t]
\plotone{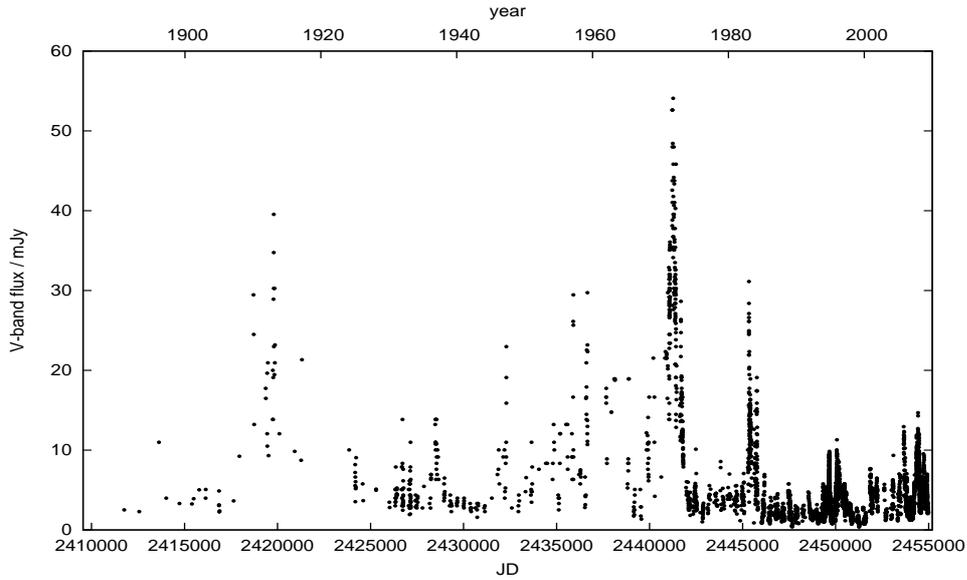}
\caption{The observations of the brightness of OJ287 from late 1800's until today.\label{fig1}}
\end{figure}
The system is not strictly periodic, but there is a simple mathematical rule which gives all major outbursts of the historical record. To define the rule, one assumes that a companion orbits the primary in a Keplerian orbit. Then demand that an outburst is produced every time the companion passes through a constant phase angle relative to the primary, and that also another outburst takes place at the opposite phase angle. Due to the nature of Keplerian orbits, this rule cannot be written in a closed mathematical form, but its consequences are easily calculated. One of the consequences is that two outburst peaks arise per period. By choosing an optimal value of eccentricity (which turns out to be 0.7) and by allowing the semimajor axis of the orbital ellipse to precess in forward direction at an optimal rate (which turns out to be $39^\circ\!$ per period), the whole historical outburst record of OJ287 is well reproduced. What is more important, the model is able to predict future outbursts. The prediction of the latest outburst on 13 September 2007 (Valtonen 2007,2008) was accurate to one day, leaving little doubt about the capability of the model (see Figure 2; the predicted time refers in this case to the start of the rapid flux rise, i.e. the beginning of the phase 2 of the outburst, as described in the next section).

The more advanced versions of the model make use of the Post Newtonian terms in the two-body orbit calculation as well as likely astrophysical details of the radiation mechanism of the outbursts. A natural way to induce two outbursts per orbit is to consider an accretion disk around the primary body, and to associate the two opposite phase angles with impacts of the secondary on this disk. By using the standard  $\alpha$ disk theory of Shakura and Sunyaev (1973) and its extension to magnetic disks by Sakimoto and Coroniti (1981), the problem remains mathematically well defined. In fact, using only 6 outbursts as fixed points in the orbit, it is possible to solve five orbital parameters in the first approximation (Valtonen 2007). The success of this model in predicting the 2007 outburst has encouraged us to add the remaining well defined outbursts in the model (9 in all), and to solve the problem for 8 parameters. In this paper we first describe the method of solution, and then list the orbital parameters. Finally, we show that the parameters fit very well with what we would expect of binary black holes, and show that the likelihood of finding an OJ287-like object in the survey of the whole sky is of the order of unity, i.e. we have not been particularly lucky in the discovery of this binary system in the nearly whole sky search of periodic variability of quasars which was initiated by Tuorla observatory in 1980.

As a side result, we may test the idea that the central body is actually a black hole. One of the most important characteristics of a black hole is that it must satisfy the so called no-hair theorem or theorems (Israel 1967, 1968, Carter 1970, Hawking 1971, 1972; see Misner, Thorne and Wheeler 1973). A  practical test was suggested by Thorne and Hartle (1985) and Thorne, Price and Macdonald (1986). In this test the quadrupole moment $Q$ of the spinning body is measured. If the spin of the body is $S$ and its mass is $M$, we determine the value of $q$ in
\begin{eqnarray}
Q = -q \frac{S^2}{Mc^2}.
\end{eqnarray}
For black holes $q=1$, for neutron stars and other possible bosonic structures $q > 2$ (Wex and Kopeikin 1999, Will 2008).

\section{OJ287 Outburst Structure}

Besides the quasiperiodic pattern of outbursts at about 12 year intervals, there is another crucial piece of information which helps in accurate orbit determination. The radiation at the outburst peaks is most likely thermal bremsstrahlung radiation from gas at about $10^5$ K temperature. This is evident from the optical-UV spectral energy distribution (Ciprini \& Rizzi 2008) as well as from the unpolarized nature of the excess radiation at the outbursts. In contrast, the optical radiation of OJ287 is non-thermal highly polarized synchrotron emission at 'normal' times, with the optical-UV spectral index of about -1.5. The reason for the thermal bremsstrahlung is thought to be an impact of a secondary black hole on the accretion disk of the primary (Lehto \& Valtonen 1996, Sundelius et al. 1997). An outburst begins when a bubble of gas torn off the accretion disk expands, cools and at some point becomes transparent at optical wavelengths. When such a bubble is viewed from a distance, the emission is seen to grow in a specific way as the observational front advances into the bubble. The size of the bubble, and thus also the rate of development of the outburst light curve, is a known function of distance from the center of the accretion disk. The process has three stages: (1) a spherical bubble becomes transparent at its forward side which causes the initial rise in the light curve. (2) When the bubble is seen approximately half-way through, it is fully transparent, and the flux rises sharply in the light-crossing time of the bubble. (3) The fully transparent bubble continues its expansion adiabatically, and consequently the radiation falls in power law manner. Figure 2 illustrates the three stages for the observations of the 2007 outburst, with the dash-dot and dashed lines outlining the first and the third stages of the outburst, respectively. From the point of view of timing, it is important to identify the initial moment of transparency when stage (1) begins. For the 2007 outburst this moment was at late hours of 10 September GMT; expressed as fractional years the moment is 2007.692 (the line drawn earlier than this has no physical significance in Figure 2). Similar timing fits have been carried for all 9 outbursts, and the corresponding timings are listed in Table 1, including the measurement uncertainties. Note that the timing ranges are slightly narrower than in Valtonen et al. (2010) where the spin determination was first carried out. It makes the finding of solutions harder but not impossible.


  \begin{figure}[t]
\includegraphics[angle=270, width=4in] {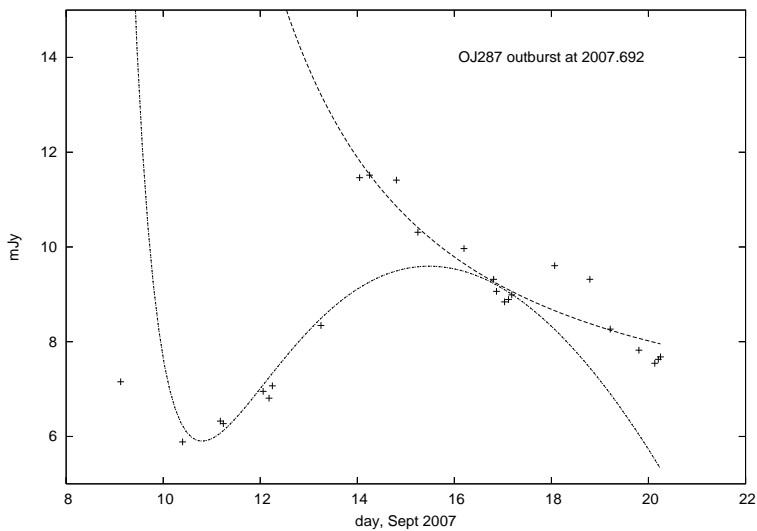}
  \caption{Observations of the 2007 outburst (crosses) together with the lines modelling the phase (1) and phase (3) evolution of the radiating bubble. Phase (2) evolution happens on 13 September when the brightness increases sharply.\label{fig2}
}
\end{figure}

\section{PN-accurate orbital description}

 We calculate the binary orbit using PN-accurate orbital dynamics that
includes the 3PN-accurate conservative non-spinning contributions,
the leading order general relativistic, classical spin-orbit and radiation reaction effects (Barker and
O'Connell 1975, Damour 1982, Kidder 1995, Mora and Will 2004).
The relevant PN-accurate equations of motion can be written schematically
as

\begin{eqnarray}
\ddot { {\vek x}} \equiv
\frac{d^2  {\vek x}} { dt^2} &=&
\ddot { {\vek x}}_{0} + \ddot { {\vek x}}_{1PN} \nonumber
+ \ddot { {\vek x}}_{SO}+\ddot { {\vek x }}_{Q}\\
&& + \ddot { {\vek x}}_{2PN} +  \ddot { {\vek x}}_{2.5PN} + \ddot { {\vek x}}_{3PN}\,, 
\end{eqnarray}
where ${\vek x} = {\vek x}_1 - {\vek x}_2 $ stands for the
center-of-mass
relative separation vector between the black holes with masses $m_1$ and $m_2$ and
$  \ddot { {\vek x}}_{0}  $ represents the Newtonian acceleration given by 
$ \ddot { {\vek x}}_{0} = -\frac{ G\, m}{ r^3 } \, {\vek x} $; $m= m_1 + m_2$ and $ r = | {\vek x} |$.
The PN contributions occurring at the conservative 1PN, 2PN, 3PN and the reactive 2.5PN orders, denoted
by $\ddot { {\vek x}}_{1PN}$, $\ddot { {\vek x}}_{2PN}$ , $\ddot { {\vek x}}_{3PN}$ and
$\ddot { {\vek x}}_{2.5PN}$ respectively, are non-spin by nature, while $ \ddot { {\vek x}}_{SO}$ is the spin-orbit term of the order 1.5PN.

The employed PN-accurate equations of motion  are in harmonic coordinates
and satify covariant spin supplementary condition. In the present analysis, we also included the precessional motion for the spin of the primary black hole due to leading order  general relativistic spin-orbit coupling.

The quadrupole-monopole interaction term $\ddot { {\vek x}}_Q $, entering at the 2PN order, reads
\begin{eqnarray}
\ddot { {\vek x}}_Q & = q \, \chi^2\,
\frac{3\, G^3\, m_1^2 m}{2\, c^4\, r^4}
\, \biggl \{
\biggl [ 5(\vek n\cdot \vek s_1)^2 
-1 \biggr ] {\vek n}
\nn
-2(\vek n\cdot \vek s_1) {\vek s_1} \biggr \},
\end{eqnarray}
where the parameter $q$, whose value is $1$ in general relativity, is introduced to test the black hole `no-hair' 
theorem. The Kerr parameter $\chi$ and the unit vector
${\vek s}_1$ define the spin of the primary black hole by the relation
${\vek S}_1 = G\, m_1^2 \, \chi\, {\vek s}_1/c$
and $\chi$ is allowed to take values between $0$ and $1$ in general relativity.  The unit vector {\vek n} lies along the direction of {\vek x}.

The initial orientation of the binary orbit is perpendicular to the accretion disk of the primary. The spin axis of the primary black hole is tilted 6 degrees relative to the spin axis of the accretion disk, and remains at this angle while the black hole spin axis precesses around the disk spin axis, with a period of 1300 yrs. The disk itself also wobbles, but with a smaller amplitude and with a period of 120 yrs. This shows up as modulation in the long term evolution of the light curve. For details, see Valtonen et al. (2009, 2010).
\begin{table}[t]
\caption{Outburst times with  estimated uncertainties. The 1982/3 outburst is used as a reference time.\label{outburst}}
\begin{tabular}{lr}
1912.970& $\pm$  0.010\\
1947.282& $\pm$  0.0005\\
1957.080& $\pm$  0.020\\
1972.940& $\pm$  0.005\\
1982.964&$\pm$  0.0\\
1984.130&$\pm$  0.002\\
1995.841&$\pm$  0.0005\\
2005.740& $\pm$  0.005\\
2007.692&$\pm$  0.0005\\
\end{tabular}
\end{table}

\section{Timing experiments}

We now require that the correct orbit must reproduce the nine outburst times of Table 1 within the stated error limits. An automatic search algorithm has been constructed which after many iterations finds a solution, starting from given initial values. The algorithm performs orbit integrations over and over again until the correct orbit is found. Because of the tolerance in the observed outburst times, the parameters of the solutions also have their associated ranges. Table 2 lists these parameters and their ranges. The latter are the three sigma error limits of the orbital parameters.

\begin{table}[h]
\caption{Solution parameters.\label{solutiontable}}
\begin{tabular}{l|r}
$\Delta \phi$  & $39^\circ\!.1 \pm 0^\circ\!.1$\\
$m_1$ & $(1.84\pm 0.01)\cdot 10^{10} M_\odot$ \\
$m_2$ & $(1.46\pm 0.1)\cdot 10^8 M_\odot$\\
$\chi$ &  $0.28\pm 0.03$\\
$\phi_0$ & $ 56^\circ\!.3\ \pm 1^\circ\!.0$\\
$e_0 $ & $0.658\pm 0.001$\\
$q$ & $1.0\pm 0.9$\\
$t_d$& $0.74\pm 0.04$\\
\end{tabular}
\end{table}

We will now discuss the physical significance of some of the parameters. The precession rate  $39^\circ\!.1$ per period is defined as the average change of the apocenter phase angle over the last 200 yrs. During each orbit the time and the phase angle is noted down when the two bodies are at their maximum separation. The exact values are determined by extrapolation between integration steps. The value given in Table 2 is the average increase of the phase angle during the 200 yr of integration, starting in 1856. A good first estimate is given by the timing of the 2005 outburst which came a year earlier than expected in non-precessing models. The precession rate leads to the value of the primary mass $m_1$ = $1.84\cdot 10^{10} M_\odot$ which is quite consistent with the values of quasar black hole masses in a large sample of SDSS quasars; the masses extend well beyond $2\cdot 10^{10} M_\odot$ (Vestergaard et al. 2008).

As to the expected frequency of such high mass values in a magnitude limited quasar sample, we may make the following estimate. OJ287 is highly variable mostly due to strong beaming. At its faintest it goes to $m_B=18$ which may be taken as its intrinsic (unbeamed) brightness. There are about $2\cdot 10^{4}$ quasars in the sky brighter than this magnitude limit (Arp 1981). Many of these quasars host binary black holes, perhaps as many as 50\% (Comerford et al. 2009). Thus potentially there are about $10^{4}$ bright binary quasars in the sky to be discovered. Vestergaard and Kelly (private communication) estimate that there is a $25\%$ probability that about one hundred of them are more massive than  $10^{10} M_\odot$. It is not unlikely that one of them would possess a jet which points more a less directly toward us, and that therefore the quasar would appear as a blazar. The brightness of the quasar is magnified by orders of magnitude by the jet, and so is the probability of its detection.

On the other hand, most of the binary black hole systems should have longer periods than OJ287 which is in the last stages of inspiral, with only $10^4$ yr left out of its potential $10^7$ yr lifetime (Volonteri et al. 2009). This compensates at least partly for the increased discovery rate due to beaming. Thus the number of bright short period binaries  in the $10^{10} M_\odot$ category which could be discovered by techniques similar to the discovery of OJ287 is of the order of unity. We may have been lucky to discover one such example, OJ287, but anyhow this is only an order of magnitude calculation, and more OJ287's may turn up in future searches of periodically varying quasars.

What about the secondary mass? Its value was calculated by Lehto and Valtonen (1996) as $m_2\approx 1.4\cdot 10^{8} M_\odot$, updated to today's Hubble constant and to the outburst peak brightness observed in 2007. This estimate is based on the astrophysics of the disk impacts and on the amount of radiation produced in these events, and thus it is totally independent of the orbit model. This agrees well with our more exact value $m_2 = 1.46\cdot 10^8 M_\odot$. The fact that the mass ratio is as high as 126 is quite natural since equal mass binaries do not easily merge within the Hubble time (Makino \& Funato 2004) and also because the accretion disk would be unstable if the mass ratio were less than 100.

The initial apocenter eccentricity at the beginning of our simulation is $e_0= 0.66$. It refers to the osculating Keplerian orbit and the osculating eccentricity value varies considerably over the orbit. If defined by using pericenter and apocenter distances as for a Keplerian orbit, the eccentricity is $e=0.7$. The initial eccentricity at the beginning of the final inspiral, say, $10^5$ yr prior to merger, must have been $e \approx 0.9$, a reasonable value at the initial stage of merging black holes (Aarseth 2008). Then by today the eccentricity would have evolved to its current value. This is typical behaviour of binary black holes of large mass ratios in galactic nuclei.

The parameter $t_d$ gives the ratio of the viscosity parameter $\alpha_g$ to the accretion rate in Eddington units. The value $t_d=0.74$ implies $14\pm0.5$ for this ratio. The mass accretion rate may be about  $0.005$ of the Eddington rate (Bassani et al. 1983) which gives $\alpha_g=0.07\pm0.03$, a reasonable value.
\begin{figure}[t] 
\plotone{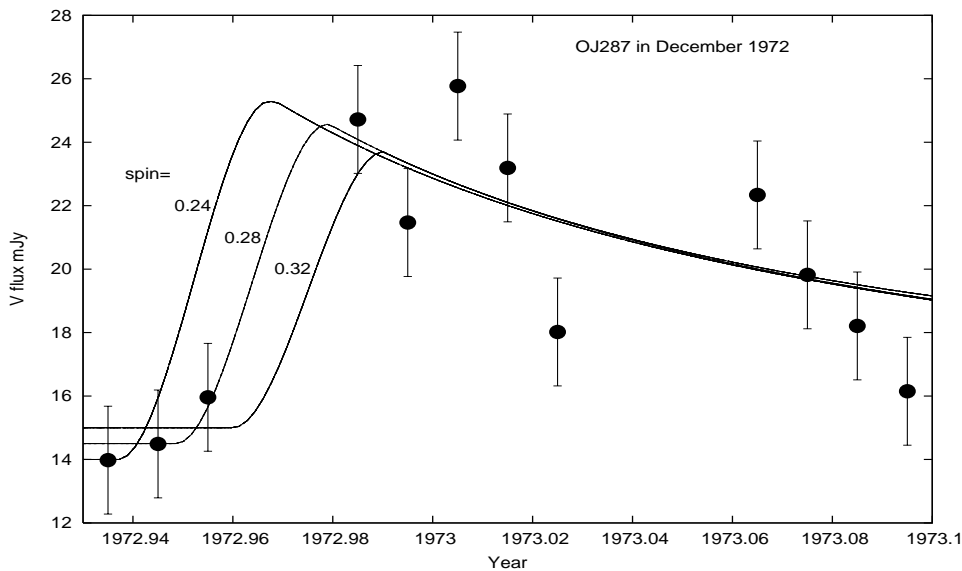}
\caption{The brightness measurements of OJ287 in 1972/3 (dots with error bars). The error bars represent short-time variability, to lesser extent measurement errors. The three lines show theoretical light curves for three values of $\chi$. \label{fig3}}
\end{figure} 

The dimensionless spin parameter is $\chi=0.28$. Some of the outburst timings are particularly sensitive to $\chi$, among them the 1973 outburst. Figure 3 illustrates how the outburst light curves depend on $\chi$ in this case. Thus we infer that the primary black hole spins approximately at one quarter of the maximum spin rate allowed in general relativity. There is an additional observation which supports this spin value. OJ287 has a basic 46$\pm$3 day periodicity (Wu et al. 2006), which may be related to the innermost stable orbit in the accretion disk. However, since we presumably observe the accretion disk almost face on, and there is an $m=2$ mode wave disturbance in the disk, this is likely to refer to one half of the period. Considering also the redshift of the system, and the primary mass value given in Table 2, this corresponds to the spin of $\chi= 0.35 \pm0.06$ (McClintock et al. 2006). For comparison, it has been estimated that $\chi\sim 0.5$ for the black hole in the Galactic center (Genzel et al. 2003). We will have a unique opportunity to improve the spin determination in 2015. If the spin is $\chi= 0.28$ the next outburst begins on 15 December, 2015. The range of possible outburst times extends from 26 November if $\chi=0.31$ to 3 January 2016 if $\chi=0.25$.
\begin{figure}[t] 
\includegraphics[angle=270, width=4in] {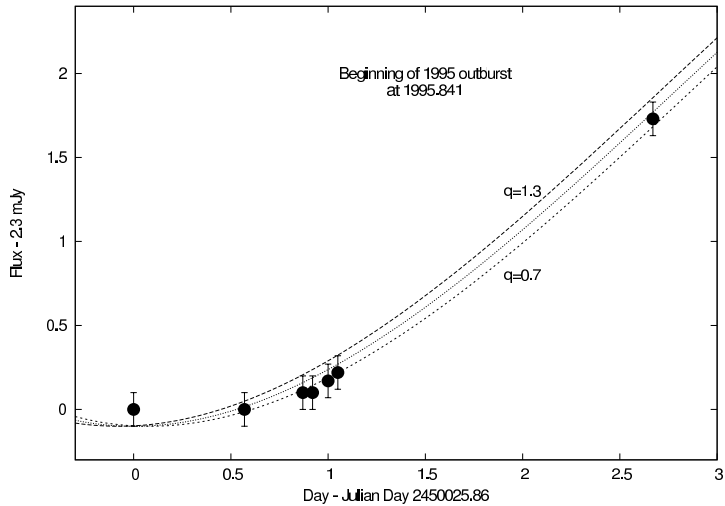}
\caption{Observations of brightness of OJ287 in 1995 (dots with error bars). The three lines are theoretical curves for the rising part of the flux (for  $q=0.7, 1.0, 1.3$).\label{fig4}}
\end{figure}

The values of $q$ cluster around $q=1.0$ with a standard deviation of $0.3$. Figure 4 illustrates the 1995 outburst and how its timing depends on $q$. The error limits of $q$ may be further narrowed down in 2019, to the level of $20\%$ (1 sigma) if the timing is carried out accurately. The fact that $q=1$ with the current accuracy indicates that the 'no-hair' theorem is valid. It is interesting that our result converges at the proper value for a general relativistic black hole.

We have also tested the sensitivity of the solution to the 2.5PN (radiation reaction) term. One may multiply the 2.5PN term by $1.0\pm0.06$ and still find a solution. Thus we confirm general relativity with $6\%$ (3 sigma) accuracy. It is possible to improve this accuracy also by a factor of 2 in the timing of the 2019 outburst. The main observational problem with this outburst is that it occurs close to the conjunction of OJ287 with the Sun. The separation in the sky is only $12^{\circ}$ at the start of the outburst on 21 July 2019, and decreases to $8^{\circ}$ during the week of the main burst. It is an observational challenge, probably too much of a challenge for ground based observations. Thus a space born telescope measurement would be desirable. This is a unique opportunity to test the 'no-hair' theorem and the radiation reaction term which will not be repeated for many decades.

\end{article}
\end{document}